\documentclass{osa-article}

\journal{oe}
\usepackage{epstopdf}
\usepackage{amsmath}

\begin{document}

%%%%%%%%%%%%%%%%%% title page information %%%%%%%%%%%%%%%%%%
\title{Beam pointing stabilization of an acousto-optic modulator with thermal control}

\author{Xiao Zhang,\authormark{1} Yang Chen,\authormark{1} Jianxiong Fang,\authormark{1} Tishuo Wang,\authormark{1} Jiaming Li,\authormark{1,3} and Le Luo\authormark{1,2,4}}

\address{\authormark{1}School of Physics and Astronomy, Sun Yat-Sen University, Zhuhai, Guangdong, China 519082\\
	\authormark{2}Department of Physics, Indiana University-Purdue University
	Indianapolis (IUPUI), Indianapolis, Indiana 46202, USA\\
	\authormark{3}lijiam29@mail.sysu.edu.cn\\
	\authormark{4}luole5@mail.sysu.edu.cn}

%\email{\authormark{*}lijiam29@mail.sysu.edu.cn} %% email address is required

% \homepage{http:...} %% author's URL, if desired

%%%%%%%%%%%%%%%%%%% abstract and OCIS codes %%%%%%%%%%%%%%%%
%% [use \begin{abstract*}...\end{abstract*} if exempt from copyright]

\begin{abstract}
Diffraction beams generated by an acousto-optic modulator (AOM) are
widely used in various optical experiments, some of which require
high angular stability with the temporal modulation of optical
power. Usually, it is difficult to realize both angular stability and
high-power modulation in a passive setup without a servo system of
radio-frequency compensation. Here, we present a method to suppress
the angular drift and pointing noise only with the thermal
management of the AOM crystal. We analyze the dependence of the angular
drift on the refractive index variation, and find that the angular
drift is very sensitivity to the temperature gradient which could
induce the refractive index gradient inside the AOM crystal. It reminds
us such angular drift could be significantly suppressed by carefully
overlapping the zero temperature gradient area with the position of the
acousto-optic interaction zone. We implement a water-cooling setup, and find that the
angular drift of an AOM is reduced over 100 times during the thermal
transient, and the angular noise is also suppressed to 1/3 of the
non-cooled case. It should be emphasized that this thermal control method is a general to suppress the beam drift in both the diffraction and the perpendicular-to-diffraction directions. The refractive index thermal coefficient of
tellurium dioxide crystal at 1064 nm determined by this angular
drift-temperature model is 16$\times$10$^{-6}$ K$^{-1}$ consistent
with previous studies. This thermal control technique provides potential applications for
optical trapping and remote sensoring that demand for intensity
ramps.
\end{abstract}

%\ocis{(000.0000) General.} % REPLACE WITH CORRECT OCIS CODES FOR YOUR ARTICLE, MINIMUM OF TWO; Avoid using the OCIS codes for “General” or “General science” whenever possible.

%%%%%%%%%%%%%%%%%%%%%%%%%%  body  %%%%%%%%%%%%%%%%%%%%%%%%%%
\section{Introduction}
Acousto-optic modulators are widely used to precisely control light
intensity and frequency, in which a traveling acoustic wave,
generated by a piezoelectric transducer, creates a modulated
refractive index and results in the Bragg diffraction of the
incoming light~\cite{Chang76}. In ultracold gas experiments, AOMs
are often applied to tune the light intensity of a high power laser
to generate an optical dipole trap (ODT) with time-dependent trap
depth for evaporative cooling or parametric cooling
~\cite{Grimm00,Li16}. These applications demand for minimizing the pointing noise of the diffracted beam so that the
wide-band pointing noise will not parametrically heat ultracold
atoms~\cite{Li16,Luo08}. Meanwhile, slow
angular drifting should also be avoided, which causes the instability of optical trapping,
especially for a cross-beam ODT~\cite{Adams95}. Previously, a large
thermal effect on the diffracted beam, induced by radio-frequency
(RF) driving, has been observed in many ODT
experiments~\cite{Kobayashi06,Frohlich07,Piggot10,Mcgehee15}. It has 
also been verified that the angular drift problem is more serious for the
high-power applications, companying with a larger AOM crystal as well as
higher RF driving power~\cite{Kobayashi06}.

There have been several methods to reduce the angular drift, such as
keeping the total RF power constant with multiple-frequency
components during thermal transient~\cite{Frohlich07,Luo08}, passing
AOM twice to cancel out the angular drift~\cite{Kobayashi06}, RF
servo schemes~\cite{Mcgehee15, Balakshy96} et. al. However, there
are still some drawbacks existing in these methods, such as
constant RF driving at high power causing the beam shape into
elliptical~\cite{note1}, double-passing setup decreasing 25$\%$ of the output
power. Especially, most of the above methods can not compensate the angular drift along the perpendicular axis
of the diffraction direction, for example Piggott detected a 0.1
mrad angular drift of CrystalTech 3080-197 AOM in the vertical
direction~\cite{Piggot10}.

In this paper, we theoretically modeled the temperature distribution in the undiffractive direction and measured the refractive index thermal coefficient of tellurium dioxide crystal. We then proposed and demonstrated a method to reduce the angular drift by water-cooling. The angular drift
of an AOM has been reduced over 100 times during the thermal transient
comparing with the non-cooling case, and the angular noise is also
suppressed to 1/3 accordingly. Furthermore, we also find this thermal control method is general to the angular drift suppression in the diffraction direction.

\section{Experimental setup and theoretical model}

\begin{figure}[ht]
\includegraphics[width=\columnwidth]{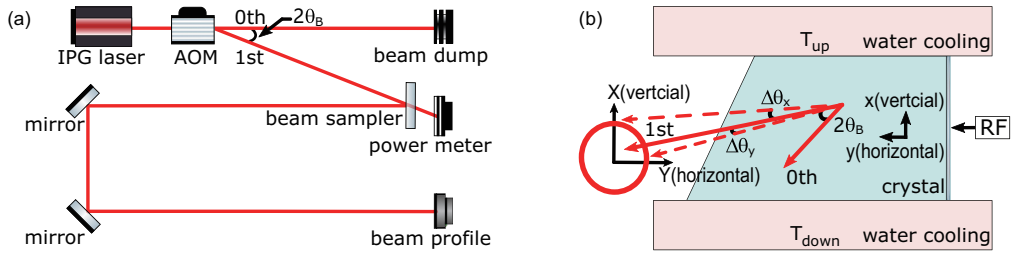}
\caption{Measurement of the angular drift of the diffracted beam of
an AOM. (a) Experimental setup. Light source is a 1064 nm CW fiber
laser (IPG, YLR-100-1064-LP) with an output power of 12 W. The
Gaussian Beam waist is 1.1 mm. Beam sampler (Thorlabs, BSF10-B)
picks out 2\% of the laser power. A CMOS beam profiler (Dataray,
S-WCD-LCM4C) with a 2048(H)$\times$2048(V) pixels and 5.5 $\times$
5.5 $\mu$m pixel size is put at $D$=9.5 meter away from the AOM. (b) AOM
structure. T$_{\text{up}}$ and T$_{\text{down}}$ are the temperature
of the top and bottom water-cooling plates respectively. They are
controlled by two chilled systems independently. \label{Fig1}}
\end{figure}

The schematic of the angular drift $\Delta\theta$ measurement is shown in
Fig.~\ref{Fig1}a. The separation angle $2\theta_B$ between the zeroth
and first order diffracted beams is determined by $\lambda f/V$, where
$\lambda$ is the wavelength of the incoming light, $f$ is the
radio-frequency (RF) of the acoustic wave, and $V$ is the
acoustic-velocity of the medium. The first order diffraction beam is
picked and measured by a beam profiler at the far end. The exposure
time of the beam profiler is limited to sub-millisecond with the
sampling rate of 10 Hz. We fit the beam profile in the horizontal
and vertical direction with a 2D Gaussian profile, and use the
center of the Gaussian fitting as the beam position $(X,Y)$. Therefore, it is easy to get $\Delta\theta_{x(y)}=\Delta$X(Y)$/D$, where $\Delta$X(Y) is the position shift related to the initial position at the far end. The AOM (IntraAction, ATM-804DA6B, tellurium dioxide) has an aperture size of 3 mm. A cartoon structure of the AOM is illustrated
in Fig.~\ref{Fig1}b. AOM is driven by a 80 MHz RF wave along the
horizontal direction. The highest diffraction efficiency of the
first order is about 92\% at 4.5 Watt RF power. The Bragg
diffraction angle is around 10 mrad at 1064 nm wavelength.
T$_{\text{up}}$ and T$_{\text{down}}$ are measured by two thermal couples from
15 to 25 \textcelsius $\ $ with 0.1 \textcelsius $\ $precision.

\begin{figure}[ht]
\includegraphics[width=\columnwidth]{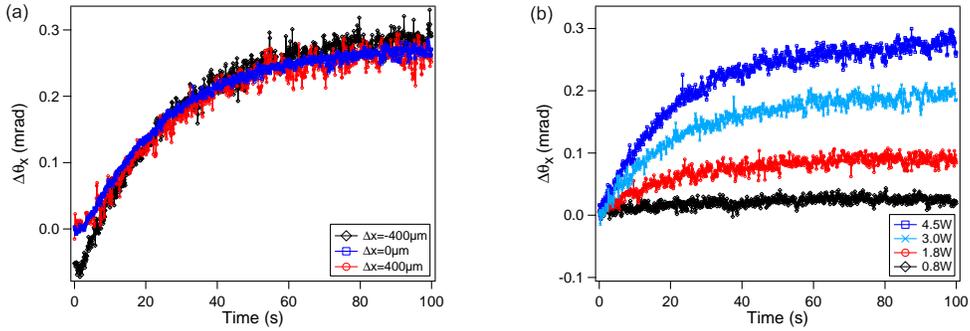}
\caption{AOM angular drift due to thermal transient. (a)
$\Delta\theta$ in the vertical direction are measured with different
input locations of the input light beam. $\Delta$X is the off center distance of the laser injection position at AOM in the perpendicular-to-diffraction direction. (b) $\Delta\theta$ for different RF driving
power with the incoming light location at $\Delta x=0$ (middle point). Blue
square, light blue cross, red circle, and black diamond are measured
under 4.5, 3.0, 1.8, and 0.8 W RF power, respectively. \label{Fig2}}
\end{figure}

We measured $\Delta\theta$ of the AOM caused by thermal transients,
as shown in Fig.~\ref{Fig2}. For a non-cooled AOM, the beam position
in vertical direction is slowly drifted to one side in about
10 s. The magnitude of $\Delta\theta$ is almost linear proportion to
the RF driving power. With the increase of the RF power,
T$_{\text{down}}$ increases slower than T$_{\text{up}}$
because a copper mount is used in the down side to mount the AOM.

The temperature space-temporal distribution of the AOM crystal
$\Delta T(x,t)$ can be approximated by a 1D heat diffusion equation
~\cite{Sheldon82}
\begin{equation}
\frac{\partial^2 T(x,t)}{\partial x^2}+\frac{\dot{q}}{k}=\frac{1}{\alpha }\frac{\partial T(x,t) }{\partial t}
\label{eqn:1}
\end{equation}
The RF driving generates an uniform energy per unit volume $\dot{q}$
inside the AOM. $\alpha$ is the thermal diffusivity, and $k$ is the
thermal conductivity. When $t\rightarrow \infty$, the system reaches
to a steady state. The temperature distribution along the $x$
direction is~\cite{Incropera}
\begin{equation}
T(x,\infty)=\frac{\dot{q}L^2}{2
k}(1-\frac{x^2}{L^2})+\frac{T_{\text{up}}-T_{\text{down}}}{2}\frac{x}{L}+\frac{T_{\text{up}}+T_{\text{down}}}{2}
\label{eqn:2}
\end{equation}
where $L$ is the height of the crystal.

When the RF driving is turned on, the thermal transient inside the
crystal can be simplified as
\begin{equation}
T(x,t)=\theta_i e^{-t/\tau}+ T(x,\infty)
\label{eqn:3}
\end{equation}
where $\theta_i\equiv T(x,0)-T(x,\infty)$ is the initial temperature
difference. The temperature difference $\Delta T(x,t)=T(x,t)-T(x,0)$
affects the refraction of the crystal. The refractive index as a
function of radius and time can be obtained from $\Delta T(x,t)$ as
expressed
\begin{equation}
n(x,t)=n_0+\frac{dn}{dT}\Delta T(x,t)
\label{eqn:4}
\end{equation}
where $n_0$ is the refractive index at the initial temperature. This
equation assumes a decrease in refractive index with increased
temperature.

\begin{figure}[ht]
\includegraphics[width=0.4\columnwidth]{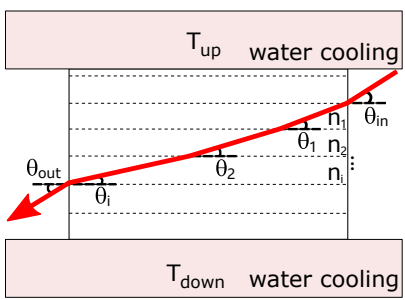}
\centering \caption{Schematic of the angular drift due to refraction
distribution $n(x,t)$.  \label{Fig3}}
\end{figure}

Fig.~\ref{Fig3} illustrates the angular drift model for the
refraction gradient. We simplify the practical case by assuming 1D
temperature gradient, so that $n_{\text{i}} \cos \theta_{\text{i}} =
n_{\text{i+1}}\cos\theta_{\text{i+1}}, i\in{1,2,3\ldots}$, and
results in the relation between the input and output angle
\begin{equation}
\sin^2\theta_{\text{out}}-\sin^2\theta_{\text{in}}=n_i^2-n_0^2
\label{eqn:5}
\end{equation}
With small angle approximation, the angular shift is given by
$\Delta\theta=\theta_{\text{out}}-\theta_{\text{in}}$
\begin{equation}
\Delta\theta= \frac{dn}{dT} \frac{dT}{dx} n_0 W
\label{eqn:6}
\end{equation}
where $W$ is the length of the crystal.

It is notable that we ignore the beam displacement effect at the AOM in Eq.~\ref{eqn:6}. The reason is that, when $W\ll D$, the position shift inside the crystal is much smaller than the position shift after the AOM. We confirm this by measuring the angular shift at different values of $D$. The result turns out the beam displacement effect is too small to be considered. 

Eq.~\ref{eqn:2} and Eq.~\ref{eqn:6} quantitatively explain the drift
effect shown in Fig.~\ref{Fig2}. Without water cooling,
$T_{\text{up}}>T_{\text{down}}$, results in a temperature gradient
along the vertical direction which causes $\Delta\theta$ with RF
ramps. In such case, there is no zero temperature gradient area for
suppressing $\Delta\theta$.

\section{Results}

\subsection{Refractive index thermal coefficient determination}
\begin{figure}[ht]
\includegraphics[width=0.6\columnwidth]{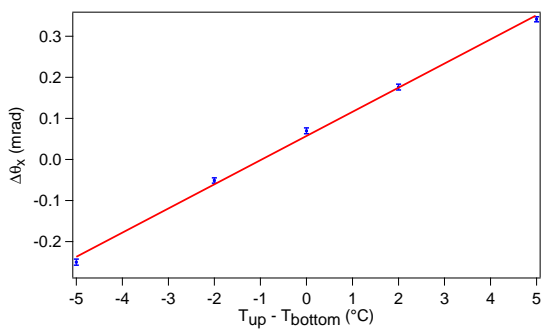}
\centering \caption{The dependence of $\Delta\theta$ on the
temperature difference between the top and bottom surfaces of an
AOM. The linear fitting result is $\Delta\theta/\Delta T = 0.059$ mrad/\textcelsius. \label{Fig4}}
\end{figure}

We fix $T_{\text{bottom}}$ to 20 \textcelsius$\ $and change $T_{up}$
from 15 to 25 \textcelsius. Then the steady angular drift of the
beam is measured with a constant RF power. The result is presented
in Fig.~\ref{Fig4}. The linear dependence of $\Delta\theta$ on the
temperature difference supports the theoretical model.

In this static measurement, $\dot{q}L^2/2 k$ can be ignored. We simulate
$dT/dx$ with Eq.~\ref{eqn:2}, and get $dT/dx\approx\Delta T/2L$, which means the temperature gradient is constant.
From Eq.~\ref{eqn:6}, we can easily obtain $dn/dT=16\times10^{-6}/K$ with $n_0=2.2079$ at 20 \textcelsius~\cite{Uchida71}, $W$=20 mm, $L$=6 mm, which is closed
to previous results~\cite{Li07}.

\subsection{Reduce the angular drift}
\begin{figure}[ht]
\includegraphics[width=\columnwidth]{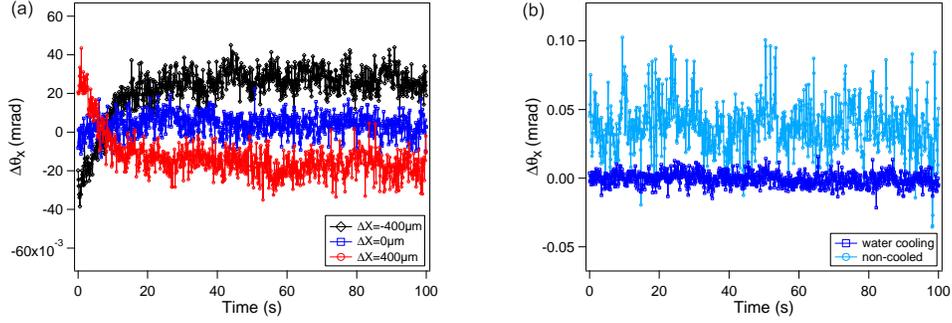}
\caption{AOM angular drift of the thermal transient with a water
cooling setup. (a), the position drift in the vertical direction of
AOM, when the laser position is in different value of the AOM. (b),
steady position of (a) and position noise comparison between water
cooling and non-cooled cases. The standard derivations of
water-cooled experiments is 5.5 $\mu$rad, while the non-cooled one
is 17.5 $\mu$rad.  \label{Fig5}}
\end{figure}

In order to minimize the angular drift of the AOM, we operate the
experiments with water cooling, and $T_{up}=T_{down}=20$
\textcelsius. From Eq.~\ref{Fig2}, we can get a parabolic-like
temperature distribution with the highest temperature and zero
temperature gradient in the middle point, resulting in null $dn/dx$ in
this spot as shown in Fig.~\ref{Fig5}, the angular shift is reduced from 0.60 mrad to zero mrad level
from non-cooling case to cooling-case. From Fig.~\ref{Fig5}a and
Fig.~\ref{Fig2}a, we can estimate that the magnitude of the
refractive index gradient in the middle point is reduced about 10
times by water-cooling. Fig.~\ref{Fig5}b shows that the water-cooled
AOM has 3 times lower angular noise than the non-cooled one, which
could be explained by weakening the heating of the
crystal~\cite{Balakshy96}. Using the water cooling scheme, our AOM
system could be suitable for cooling in micro-ODT ~\cite{Wenz13} and
lattice trap~\cite{Blatt15}, both of which highly demand pointing
stability and noise reduction.

\begin{figure}[ht]
\includegraphics[width=0.5\columnwidth]{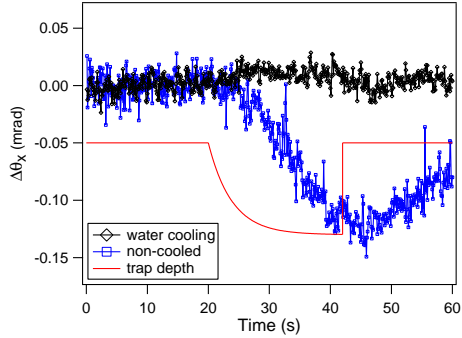}
\centering \caption{The comparison of the angular drift during an
evaporative cooling process between water-cooled and non-cooled AOMs
with exponential decay RF power for evaporative cooling. The trap depth is lower to 1$\%$ in 20 seconds. \label{Fig6}}
\end{figure}

We test the evaporative cooling with the water-cooled AOM~\cite{Ohara01}. The result is shown in Fig.~\ref{Fig6}, indicating that the cooling setup
could solve the angular drift problem fairly well.

\begin{figure}[ht]
\includegraphics[width=0.5\columnwidth]{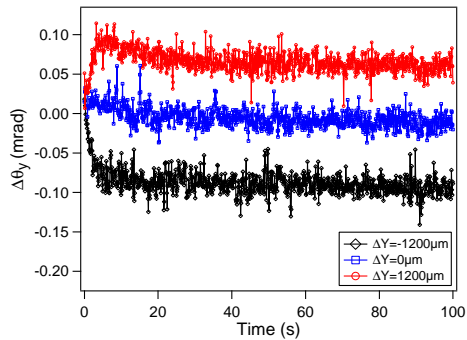}
\centering \caption{Measurement results of angular drift in the Bragg diffraction direction with a water cooling setup in the opposite side of piezoelectric transducer. $\Delta$Y is the off-the-center distance of the laser injection position at AOM in the diffraction direction.\label{Fig7}}
\end{figure}

We also use the cooling scheme to suppress the angular drift along
the diffraction direction. Different with the perpendicular
direction, there is an attached piezoelectric transducer at one side
of the diffracting direction, which can not be water-cooled. The
solution is that we can control the temperature of the other side to the same temperature as the transducer side  by thermal control to
make an symmetry temperature distribution along the diffraction
direction~\cite{Kobayashi06}. The result is shown in
Fig.~\ref{Fig7}.

\section{Conclusions}
We studied the angular drift and noise of the AOM with RF power
ramps. A thermal-control setup is applied to reduce the angular
drift significantly as well as the pointing-noise.  The results
agree with the the refractive index-temperature model of the AOM
crystal very well. This technique opens a route for high-precision
beam control for laser cooling and trapping with micro-ODT and
lattice trap.

\section*{Funding}
National Natural Science Foundation of China (NSFC) under Grant No.11804406, Fundamental Research Funds for Sun Yat-sen University 18lgpy78. Science and Technology Program of Guangzhou 2019-03-01-05-3001-0035. NSFC No.11774436. SunYat-sen University Discipline Construction Fund, Guangdong ProvinceYouth Talent Program under Grant No.2017GC010656.

%%%%%%%%%%%%%%%%%%%%%%% References %%%%%%%%%%%%%%%%%%%%%%%%%
% \begin{thebibliography}{1}
\bibliography{reference}

\begin{thebibliography}{10}
\bibitem{note1}
\enquote{We measured three aoms, such as intraaction model atm-804da6b, isomet
  model m1306-t80l-4 and model m1135-t80l-3. all of their diffraction beams
  become elliptical when the high driving power reaches half of the maximum
  diffraction efficient,} .
\end{thebibliography}
%\begin{thebibliography}{99}
\newcommand{\enquote}[1]{``#1''}

\end{document}